\documentclass[a4paper,reqno,12pt,draft]{article}
\usepackage{amssymb,euscript,bbold}
\newcommand{\be}{\begin{equation}}
\newcommand{\ee}{\end{equation}}
\newcommand{\ba}{\begin{eqnarray}}
\newcommand{\ea}{\end{eqnarray}}

\newcommand{\ft}{\footnote}

\newcommand{\we}{\wedge}

\begin{document}
\input{epsf}

\begin{flushright}
RUNHETC-2000-44
\end{flushright}
\begin{flushright}
\end{flushright}
\begin{center}
\Large{\sc On realising N=1 Super Yang-Mills in $M$ theory.}\\
\bigskip
{\sc B.S. Acharya}\ft{bacharya@physics.rutgers.edu}\\
\smallskip\large
{\sf Department of Physics,\\
Rutgers University,\\ 126 Freylinghuysen Road,\\ NJ 08854-0849.}

\end{center}
\bigskip
\begin{center}
{\bf {\sc Abstract}}
\end{center}
Pure ${\cal N}$ =1 super Yang-Mills theory can be realised as a
certain low energy limit of $M$ theory near certain singularities
in $G_2$-holonomy spaces. For $SU(n)$ and $SO(2n)$ gauge groups these
$M$ theory backgrounds can be regarded as strong coupling limits
of wrapped D6-brane configurations in Type IIA theory on certain
non-compact Calabi-Yau spaces such as the deformed conifold.
Various aspects of such realisations are
studied including the generation of the superpotential, domain walls,
QCD strings and the relation to recent work of Vafa. In the spirit of
this recent work
we propose a `gravity dual' of $M$ theory near these singularities.

\newpage

\Large
\noindent
{\bf {\sf 1. Introduction.}}
\normalsize
\bigskip

Pure super Yang-Mills theory in four dimensions is an interesting gauge theory
where we believe we can say something about the physics at strong coupling.
The model shares some of the features of ordinary QCD, but is easier to
understand because of the added feature of supersymmetry. 

This super Yang-Mills theory can be realised in various limits of string theory
and $M$ theory vacua. These are typically limits of the world-volume theories
on intersecting or wrapped branes or limits of theories localised near
singularities \cite{oog,wit,kv,b1,polstras,klebstras,malnun,vafa}\ft{
For a review of the approach via branes with many more references see
\cite{giv}}. 
As we will see, some of these studies are closely related.

In \cite{b1} we studied some aspects of $M$ theory physics near
certain singularities in $G_2$ holonomy spaces. Our motivation for this
was to begin to answer the question: what is the physics of $M$ theory
compactified on a $G_2$-holonomy space? The motivation for the question
was that its answer might eventually shed some new light on the physics
of four dimensional theories with ${\cal N}$=1 supersymmetry. 

The singularities in question
took the form of families of {\sf {ADE}} singularities in $\mathbb{R^4}$ 
parametrised
by a supersymmetric 3-cycle $M$. In other words, the total space of the
singular 7-manifold $J$ took the form of a singular fiber bundle with fibers
$\mathbb{{R^4}/{\Gamma}}$ and base $M$. The zero-section of this
bundle, which is a copy of $M$, is an associative 3-cycle in the
singular $G_2$-holonomy space ie a supersymmetric 3-cycle.
When the volume of $M$ is large the 7-manifold looks approximately like
flat $\mathbb{{R^4}/{\Gamma}}{\times}\mathbb{R^3}$ 
and $M$ theory on $J{\times}\mathbb{R^{3,1}}$
looks approximately like $M$ theory on $X$ $\equiv$
$\mathbb{{R^4}/{\Gamma}}{\times}\mathbb{R^{6,1}}$.
When $\mathbb{\Gamma}$ is a finite {\sf {ADE}} subgroup of $SU(2)$ acting on
$\mathbb{C^2}$ $\equiv$ $\mathbb{R^4}$, 
the low energy physics of $M$ theory on $X$ is described
by super Yang-Mills theory on ${0}{\times}\mathbb{R^{6,1}}$, 
the singular subset of 
$X$. We argued that the theory on $J$ should be regarded as a twisted
compactification of seven dimensional super Yang-Mills formulated on
${M}{\times}\mathbb{R^{3,1}}$. 
The massless spectrum of the effective four dimensional
theory was determined to be ${\cal N}$ = $1$ super Yang-Mills with 
${b_1}(M)$ adjoint chiral multiplets. In particular when ${b_1}(M)$ is zero
the low energy physics of $M$ theory near singularities of this 
type corresponds
to pure super Yang-Mills. To be more precise, the super Yang-Mills description
is valid when the volume of $M$ is large 
compared to the Planck scale and one considers energy scales
below the mass of the lowest Kaluza-Klein excitation.
We argued that a non-trivial superpotential in this
model is generated by fractional $M$2-brane instantons which wrap $M$. This
potential agrees with the field theoretic result.
 
For
gauge group $SU(n)$ ie ${\mathbb{\Gamma}}$ $\equiv$ $\mathbb{Z_n}$
a natural proposal is that this
$M$ theory background 
is the strong string coupling limit of Type IIA string theory on
$T^{*}(M)$ with $n$ D6-branes wrapped around the zero section \cite{b1}.

The original motivation for this work was to elaborate on the
discussion in \cite{b1} by giving a stronger argument for the generation of
the superpotential and also to address issues such as domain walls and QCD
strings.
We will partly achieve this by presenting much simpler explicit examples
of the singular $M$ theory background spacetime in which $M$ is a three sphere
$S^3$. In this case $J$ is a $\mathbb{Z_n}$-orbifold of the standard spin bundle
$S({S^3})$ of $S^3$. ${S}({S^3})$ is topologically trivial ie
$S({S^3})$ $\equiv$ $\mathbb{R^4}{\times}{S^3}$, but admits a non-trivial
bundle metric with $G_2$-holonomy
\cite{bryssal,gpp}. The quotient by $\mathbb{Z_n}$ which defines
$J$ is topologically equivalent to
$\mathbb{R^4}/{\mathbb{Z_n}}{\times}{S^3}$ where the $\mathbb{Z_n}$ acts on
each fiber as an {\sf ADE} subgroup of $SU(2)$ acting on  
$\mathbb{C^2}$ $\equiv$ $\mathbb{R^4}$. The singularities of $J$ are precisely
as described above: a family of $A_{n-1}$ singularities fibered over $S^3$.
Most importantly, $\mathbb{Z_n}$ preserves the $G_2$-structure on 
$S({S^3})$ so that $J$ has $G_2$-holonomy.
This construction can be generalised to any {\sf ADE} subgroup of $SU(2)$. In
the $SU(n)$ case however which will be our primary focus in this paper,
$M$ theory on $J$ is naturally proposed as the strong coupling limit
of Type IIA string theory on ${T^*}({S^3})$ with $n$ D6-branes wrapping the
$S^3$. This latter string theory background was recently
studied in \cite{vafa}. As we will see, combining the results of \cite{vafa}
with those presented here gives a strong indication that the proposal is
correct.

One of the central themes of \cite{vafa} was to propose a
closed string, gravity dual of the open string theory describing the
D6-branes. Given that we have proposed a strong coupling limit of the
latter and have gathered evidence in its favour, it is natural to
ponder the existence of an $M$ theory description of the gravity dual
proposed in \cite{vafa}. We will make a proposal of just such a dual here.
This $M$ theory background has some distinguishing features. For instance
there are $\mathbb{Z_n}$-charge strings which we suggest should be identified
with the super QCD-strings in the correspondence with Yang-Mills theory.
The model also contains domain walls.

This paper is organised as follows. In the next section
we aim to clarify aspects of the superpotential
in these models by compactifying the $M$ theory background on a circle
and considering Type IIA theory on $J$. Here we will make contact with some
field theory results \cite{h1} which will be reinterpreted here
in terms of fractional branes. In particular, we will see that the description
of fractional branes as wrapped branes \cite{d} is already anticipated in
quantum Yang-Mills theory. Next we make some comments concerning BPS-saturated
domain walls in these models and tentatively identify domain walls of the
expected tension. Finally, in the light of \cite{vafa}, we propose a
`gravity dual' of $M$ theory on $J$.

Whilst this work was in progress we learned of the forthcoming work of
Atiyah, Maldacena and Vafa \cite{malvaf} 
in which the same models are studied. We wish to
thank J. Maldacena for informing us of this work. We would also 
like to mention the
recent paper \cite{mir} which discusses smooth supergravity solutions
describing branes on special holonomy manifolds (including those described
in \cite{bryssal,gpp}) and is relevant for a discussion of gravity duals of
the corresponding world-volume theories.

\newpage
\Large
\noindent
{\bf {\sf 2. Type IIA theory on J.}}
\normalsize
\bigskip

Consisder Type IIA string theory compactified to three dimensions on a
seven manifold $X$ with holonomy $G_2$. If $X$ is smooth we can determine the
massless spectrum of the effective supergravity theory in three dimensions
as follows. Compactification on $X$ preserves four of the 32 supersymmetries
in ten dimensions, so the supergravity theory has three dimensional 
${\cal N}$ = 2
local supersymmetry.  The relevant bosonic fields of the ten dimensional 
supergravity theory are the metric, $B$-field, dilaton plus the Ramond-Ramond
one- and three-forms. These we will denote by $g,B,{\phi},{A_1},{A_3}$ 
respectively. 
Upon Kaluza-Klein reduction the metric gives rise to a three-metric
and ${b_3}(X)$ massless scalars. The latter parametrise the moduli space of
$G_2$-holonomy metrics on $X$. $B$ gives rise to
${b_2}(X)$ periodic scalars ${\varphi}_i$. 
$\phi$ gives a three dimensional dilaton. $A_1$ reduces to
a massless vector, while $A_3$ gives ${b_2}(X)$ vectors and 
${b_3}(X)$ massless
scalars. 
In three dimensions a vector is dual to a periodic scalar, so 
at a point in moduli space where the vectors are free we can dualise them.
The dual of the vector field originating from $A_1$ is the period of the
RR 7-form on $X$, whereas the duals of the vector fields coming from $A_3$
are given by the periods of the RR 5-form $A_5$ over a basis of 5-cycles
which span the fifth homology group of $X$. Denote these by scalars
by ${\sigma}_i$.
All in all, in the dualised theory
we have in addition to the supergravity multiplet, ${b_2}(X) +{b_3}(X)$ 
scalar 
multiplets. Notice that ${b_2}(X)$ 
of the scalar multiplets contain two real
scalar fields, both of which are periodic.

Now we come to studying the 
Type IIA theory on $J$. Recall that $J$ is defined as
an orbifold of the standard spin bundle of $S^3$.  
To determine the massless spectrum of IIA string theory on
$J$ we can use standard orbifold techniques. However, the answer can be phrased
in a simple way. $J$ is topologically 
$\mathbb{R^4}/\mathbb{Z_n} {\times} S^3$. This
manifold can be desingularised to give a smooth seven manifold 
$M^{\mathbb{Z_n}}$
which is topologically ${X^{\mathbb{Z_n}}}{\times}S^3$, 
where $X^{\mathbb{Z_n}}$ is
homeomorphic to an $ALE$ space.
The string theoretic cohomology groups of $J$ are isomorphic to the usual
cohomology groups of $M^{\mathbb{Z_n}}$. 
The reason for this is simple: $J$ is a global
orbifold of $S({S^3})$. The string theoretic cohomology groups count
massless string states in the orbifold CFT. The massless string states in
the twisted sectors are localised near the fixed points of the action of
$\mathbb{Z_n}$ on the spin bundle. Near the fixed points we can work on
the tangent space of $S(S^3)$ near a fixed point and the action of $\mathbb{Z_n}$
there is just its natural action on $\mathbb{R^4}{\times}\mathbb{R^3}$.

Note that blowing up $J$ to give $M^{\mathbb{Z_n}}$
cannot give a metric with $G_2$-holonomy which is continuosly connected
to the singular $G_2$-holonomy metric on $J$, since this would require
that the addition to homology in passing from $J$ to 
$M^{\mathbb{Z_n}}$ receives contributions
from four-cycles. This is necessary since these are dual to elements of
$H^3 (M)$ which generate metric deformations preserving the $G_2$-structure.
This argument does not rule out the possibility that 
$M^{\mathbb{Z_n}}$ admits `disconnected'
$G_2$-holonomy metrics, but is consistent with the fact that pure 
super Yang-Mills theory in four dimensions does not have a Coulomb branch.

The important points to note are that the twisted sectors contain massless
states consisting of $r$ scalars and $r$ vectors where $r$ is the rank of
the corresponding {\sf {ADE}} group associated to 
$\mathbb{\Gamma}$. In the case of
primary interest $r$ = $n-1$.
The $r$ scalars can 
intuitively thought of as the periods of the $B$ field through $r$ two cycles.
In fact, for a generic point in the moduli space of the orbifold conformal
field theory the spectrum contains massive particles charged under the
$r$ twisted vectors. 
These can be interpreted as wrapped D2-branes whose quantum
numbers are precisely those of $W$-bosons associated with the breaking of
an {\sf {ADE}} gauge group to $U(1)^r$. This confirms our interpretation of the
origin of this model from $M$ theory:
the values of the $r$ $B$-field scalars can be interpreted 
as the expectation values of Wilson lines around the eleventh dimension
associated with this symmetry breaking. At weak string coupling and large
$S^3$ volume these states are very massive and the extreme low energy
effective dynamics of the twisted sector states is described by ${\cal N}$=2
$U(1)^r$ super Yang-Mills in three dimensions. 
Clearly however, the underlying
conformal field theory is not valid when the $W$-bosons become massless. The
appropriate description is then the pure super Yang-Mills theory on 
$\mathbb{R^3}{\times}{S^1}$ which corresponds to a sector of $M$ theory on
$J{\times}S^1$.
In this section however, our strategy will be to 
work at a generic point in the CFT moduli space which corresponds to being
far out along the Coulomb branch of the gauge theory. 
We will attempt to calculate the
superpotential there and then continue the result to four dimensions. 
This
exactly mimics the strategy of \cite{sw,h1} in field theory and \cite{kv}
in the context of $F$-theory. 
Note that we are implicitly ignoring gravity here. More precisely, we are
assuming that in the absence of gravitational interactions with the twisted
sector, the low energy physics of the twisted sectors of the CFT
is described by the Coulomb branch of the gauge theory. This is natural since
the twisted sector states are localised at the singularities of 
$J{\times}\mathbb{R^{2,1}}$ whereas the gravity propagates in bulk.

In this approximation, 
we can dualise the photons to obtain a theory of $r$ chiral
multiplets, each of whose bosonic components ({\mbox{\boldmath ${\varphi}$}}
and {\mbox{\boldmath ${\sigma}$}})
is periodic. 
But remembering
that this theory arose from a non-Abelian one we learn that the moduli space
of classical vacua is
\be
{\cal M}_{cl} = {{\mathbb C}^r \over {{\Lambda}_{W}^{\mathbb{C}}\rtimes W_g}}
\ee

where ${\Lambda}^{\mathbb{C}}_W$ is the 
complexified weight lattice of the {\sf ADE} group and $W_g$ is
the Weyl group.

We can now ask about quantum effects. In particular is there a 
non-trivial superpotential for these chiral multiplets? In a theory with four
supercharges BPS instantons with only two chiral fermion zero modes can generate
a superpotential. Are there instantons in Type IIA theory on J ?
BPS instantons come from branes wrapping supersymmetric cycles and Type IIA
theory on a $G_2$-holonomy space can have instantons corresponding to D6-branes
wrapping the space itself or D2-brane instantons which wrap supersymmetric
3-cycles. For smooth $G_2$-holonomy manifolds these were studied in \cite{greg}.
In the case at hand the D6-branes would generate a superpotential
for the dual of the graviphoton multiplet which lives in the gravity multiplet
but since we wish to ignore gravitational physics for the moment, we will ignore
these. In any case, since $J$ is non-compact, these configurations have
infinite action. The D2-branes on the other hand are much more interesting. 
They can wrap the supersymmetric $S^3$ over which the singularities of $J$
are fibered. We can describe the dynamics of a wrapped D2-brane as follows.
At large volume, where the sphere becomes flatter and flatter the world-volume
action is just the so called `quiver gauge theory' 
described in \cite{dm}. Here
we should describe this theory not just on $S^3$ but on a
supersymmetric  $S^3$ embedded in
a space with a non-trivial $G_2$-holonomy structure. The upshot is that the
world-volume theory is in fact a cohomological field theory \cite{bsv,gm}
so we can trust it
for any volume as long as the ambient space has $G_2$-holonomy. Note
that, since we are ignoring gravity, we are implicitly ignoring higher
derivative corrections which could potentially also affect this claim.
Another crucial
point is that the $S^3$ which sits at the origin in 
$\mathbb{R^4}$ in the covering
space of $J$ is the supersymmetric cycle, and the spheres away from the origin
are not supersymmetric,
so that the BPS wrapped
D2-brane is constrained to live on the singularities of $J$. In the quiver
gauge theory, the origin is precisely the locus in moduli space at which the
single D2-brane can fractionate (according to the quiver diagram) and this
occurs by giving expectation values to the scalar fields which parametrise
the Coulomb branch which corresponds to the position of our D2-brane in
the dimensions normal to $J$. 

What contribution to the superpotential do the fractional D2-branes make?
To answer this we need to identify the configurations which possess only
two fermionic zero modes. We will not give a precise string theory argument
for this, but using the correspondence
between this string theory and field theory will identify exactly which
D-brane instantons we think are responsible for generating the superpotential.
This may sound like a strong assumption, but as we hope will become clear,
the fact that the fractional D2-branes are wrapped D4-branes is actually
anticipated by the field theory! This makes this assumption, in our opinion,
somewhat weaker and adds credence to the overall picture being presented here.

In the seminal work of \cite{d}, it was shown that the fractionally
charged D2-branes are actually D4-branes which wrap the `vanishing' 2-cycles
at the origin in $\mathbb{{R^4}/{\Gamma}}$. More precisely, each individual 
fractional D2-brane, which originates from a single D2-brane
possesses D4-brane charge, but the total configuration, since it began life
as a single D2-brane has zero D4-brane charge. The possible contributions to
the superpotential are constrained by supersymmetry and must be
given by a holomorphic function of the $r$ chiral superfields and
also of the holomorphic gauge coupling constant $\tau$ which corresponds
to the complexified volume of the $S^3$ in eleven dimensional $M$ theory. 
We have identified 
above the bosonic components of the chiral
superfields above. $\tau$ is given by
\be
\tau = \int \varphi + i C
\ee
where $\varphi$ is the $G_2$-structure defining 3-form on $J$. The period of
the $M$ theory 3-form through $S^3$ plays the role of the theta angle.

The world-volume action of a D4-brane contains the couplings
\be
L = B \we A_3 + A_5
\ee

Holomorphy dictates that there is also a term
\be
B \we \varphi
\ee

so that the combined terms are written as
\be
B \we \tau + A_5
\ee

Since the fourbranes wrap the `vanishing cycles' and the $S^3$ we see that
the contribution of the D4-brane corresponding to the $k$-th 
fractional D2 takes the form
\be
S = - \mbox{\boldmath ${\beta}_{k}.z$}
\ee
where we have defined
\be
{\mbox{\boldmath $z$}} = \tau {\mbox{\boldmath ${\varphi}$}} + 
{\mbox{\boldmath ${\sigma}$}}
\ee
and the \mbox{\boldmath ${\beta}_k$} are charge vectors. The 
$r$ complex fields {\mbox{\boldmath $z$}} are the natural holomorphic
functions upon which the superpotential will depend.

The wrapped D4-branes are the magnetic duals of the massive D2-branes which
we identified above as massive $W$-bosons. As such they are magnetic monopoles
for the original $SU(n)$ gauge symmetry. 
Their charges are therefore given
by an element of the co-root lattice of the Lie algebra and thus each of
the $r$ + 1 \mbox{\boldmath ${\beta}$}'s is a rank $r$ vector in this space. 
Choosing a basis
for this space corresponds to choosing a basis for the massless states in
the twisted sector Hilbert space which intuitively we can think of as a basis
for the cohomology groups Poincare dual to the `vanishing' 2-cycles. A natural
basis is provided by the simple co-roots of the Lie algebra of $SU(n)$, which
we denote by \mbox{\boldmath ${\alpha}^*_k$}
for $k = 1,...,r$.
This
choice is natural, since these, from the field theory point of view are
the fundamental monopole charges.

At this point it is useful to mention that the $r$
wrapped D4-branes whose magnetic charges are given by the simple co-roots of
the Lie algebra correspond in field theory to monopoles with charges
\mbox{\boldmath ${\alpha}^*_k$} and each of these is known to possess precisely
the right number of zero modes to contribute to the superpotential. Since we
have argued that in a limit of the Type IIA theory on $J$, the dynamics 
at low energies is governed
by the field theory studied in \cite{h1} it is natural to expect that these
wrapped fourbranes also contribute to the superpotential. Another striking
feature of the field theory is that these monopoles also possess a fractional
instanton number - the second Chern number of the gauge field on
$\mathbb{R^3}{\times}S^1$. These are precisely in correspondence with the
fractional D2-brane charges. Thus, in this sense, the field theory anticipates
that fractional branes are wrapped branes.

In the field theory on $\mathbb{R^3}{\times}S^1$
it is also important 
to realise that there is precisely one additional BPS state which
contributes to the superpotential. The key point is that this state, unlike
the previously discussed monopoles have dependence on the periodic direction
in spacetime. 
This state is associated with the affine
node of the Dynkin diagram. Its monopole charge is given by
\be
 - {\Sigma}_{k=1}^r  \mbox{\boldmath ${\alpha}^*_{k}$} 
\ee
and it also carries one unit of instanton number.

The action for this state is
\be
S =  {\Sigma}_{k=1}^r \mbox{\boldmath ${\alpha}^*_{k}.z$} -2\pi i \tau
\ee

Together, these $r$ + 1 BPS states can be regarded as fundamental in the sense
that all the other finite action BPS configurations can be thought of as
bound states of them.

Thus, in the correspondence with string theory it is
also natural in the same sense as alluded to above that a state with these
corresponding quantum numbers also contributes to the superpotential. It
may be regarded as a bound state of anti-D4-branes with a charge one D2-brane.
In the case of $SU(n)$ this is extremely natural, since 
the total D4/D2-brane charge of the $r$ +1 states is zero/one,
and this is precisely the charge of the D2-brane configuration on $S^3$
whose world-volume action is the quiver gauge theory for the affine
Dynkin diagram for $SU(n)$. In other words, the entire superpotential is
generated by a single D2-brane which has fractionated.

In summary, we have seen that the correspondence between the Type IIA string
theory on $J$ and the super Yang-Mills theory on 
$\mathbb{R^3}{\times}S^1$ is quite
striking. Within the context of this correspondence we considered a smooth
point in the moduli space of the perturbative Type IIA CFT, where the spectrum
matches that of the Yang-Mills theory along its Coulomb branch. On the
string theory side we concluded that the possible instanton contributions
to the superpotential are from wrapped D2-branes. Their world volume theory
is essentially topological, from which we concluded that they can fractionate.
As is well known, the fractional D2-branes are really wrapped fourbranes.
In the correspondence with field theory, the wrapped fourbranes are magnetic
monopoles, whereas the D2-branes are instantons. Thus if, these branes
generate a superpotential they correspond, in field theory to 
monopole-instantons. 
This is exactly how the field theory potential is known to be generated. 
We thus expect
that the same occurs in the string theory on $J$.

Finally, the superpotential generated by these states is of affine-Toda
type and is known to possess $n$ minima corresponding to the vacua of the
$SU(n)$ super
Yang-Mills theory on $\mathbb{R^4}$. The value of the superpotential
in each of these vacua is of the form ${e}^{2{\pi} i\tau  \over n}$.
As such it formally looks as though it was generated by fractional instantons,
and in this context fractional $M$2-brane instantons. This result holds
in the four dimensional $M$ theory limit because of holomorphy and thus
elaborates upon the result of \cite{b1}.

\bigskip
\Large
\noindent
{\bf {\sf 3. On Domain Walls in $M$ theory on $J$.}}
\normalsize
\bigskip

We now turn our attention to domain walls. Since the model has a finite number
of vacua one might expect that there are domain walls which separate them. This
is believed to be the case for the super Yang-Mills theory because that theory
has a spontaneously broken discrete symmetry. This is the quantum 
remnant of the
classical $U(1)$ $R$-symmetry. $M$ theory on $J$ (or
any other $G_2$-holonomy manifold) does not obviously have a $U(1)$ symmetry.
The classical supergravity theory does have a 
$\mathbb{Z_2}$ $R$-symmetry which acts as
$\pm$$\mathbb{1}$ on the supercharges and preserves the $G_2$-structure. This
order two symmetry is naturally identified with the 
$\mathbb{Z_2}$ symmetry of the
super Yang-Mills vacua.
Following this line of reasoning does not give us any obvious reason to expect
to see domain walls connecting the various vacua. However, the model does
appear to possess domain walls with some of
the right properties, as we will see.

As discussed in \cite{b2}, in $M$ theory compactification on a $G_2$-holonomy
manifold the BPS domain walls are basically
of two types. Firstly there are $M$5-branes which wrap supersymmetric three
cycles. Secondly there are $M$2-branes which sit at a point on the seven 
manifold. Actually, the general story is more complicated since $M$2-branes
and $M$5-branes can form bound states and one must not rule out the possibility
of bound state walls in the compactified theory. This is in line with the
fact that the central charge in the 
${\cal N}$ = 1 supersymmetry algebra in four dimensions
which represents the charge of domain walls is complex. 

In $M$
theory on $J$ the $S^3$ is a supersymmetric cycle, so an $M$5-brane wrapped
on it looks like a domain wall. Because it is BPS saturated its tension
is given by the volume
\be
T = Vol(S^3) = {1 \over g^2_{YM}}
\ee
which is identified as the gauge coupling in the correspondence with Yang-Mills.
One important aspect of the super Yang-Mills theory is that the t'Hooft
large $n$ limit is a good approximation to its physics. This means that
\be
T \approx n
\ee
and this is the expected behaviour of the domain walls in this model \cite{wit}.
The M$2$-brane domain walls do not have a tension which depends on the coupling.
However, the M$2$-M$5$-bound state wall has a tension whose square
is given by the sum of the squares of its two constituents. In the large $n$
limit the tension of this bound state is also of order $n$, so we cannot
distinguish the pure $M$5-brane from the bound state by this reasoning.
So, even though $M$ theory on $J$ does not appear to undergo discrete
symmetry breaking, BPS domain walls appear to exist with at least some of
the expected properties of their super Yang-Mills counterparts.

In the IIA theory on $J$ the wrapped $M$5-brane domain walls go over to
D$4$-branes wrapping the supersymmetric $S^3$. This is natural, since
the various vacua of the three dimensional theory are related by shifts of
the {\mbox{\boldmath ${\sigma}$}}. These latter fields are just the 
twisted RR scalars which can be thought of as periods of
the RR 5-form potential through the vanishing 5-cycles. Since these fields
couple naturally to the D4-branes we see that the fourbrane charges shift
by one unit from vacuum to vacuum. This lends further support to the claim
that the domain walls in $M$ theory on $J$ are $M$5-branes - possibly bound
to $M$2-branes.

\newpage

\Large
\noindent
{\bf {\sf 4. `Gravity' Dual of $M$ theory on $J$.}}
\normalsize
\bigskip

As we mentioned in the introduction, $M$ theory on $J$ is 
naturally proposed as the strong string
coupling limit of Type IIA theory on $T^{*}({S^3})$ with $n$ D6-branes wrapped
around the $S^3$. 

The IIA background consisting of $n$ D6-branes wrapped around the $S^3$
in the deformation of the conifold singularity $T^{*}({S^3})$ was studied
by Vafa in \cite{vafa}. In this work it was shown how open topological
string amplitudes can be used to calculate certain higher derivative
terms in the effective world-volume action on $\mathbb{R^4}$. 
Moreover a gravity
dual of this model was proposed. This was defined as Type IIA theory on
the resolution of the conifold singularity together with RR fluxes.
This smooth non-compact Calabi-Yau background can 
loosely be regarded as an
$\mathbb{R^4}$-bundle over $S^2$. As regards the fluxes it is crucial
that there are
$n$ units of RR 2-form flux through the $S^2$ corresponding to the
D6-brane charge.
An important part of \cite{vafa} was that the closed string topological
amplitudes are not modified in the presence of RR flux. The two sets
of topological amplitudes agreed if the complexified Kahler class of the
$\mathbb{P^1}$ $=$ $S^2$ on the gravity side was identified with the 
gaugino bilinear
superfield on the open string side. 
In the low energy limit of the D6-brane theory it was shown how the effective
superpotential for the super Yang-Mills theory emerged from these topological
amplitudes and on the closed string side these corresponded to world-sheet
instantons wrapping the $S^2$. In fact, on the open string side these have
the form of fractional D2-brane instantons wrapping the $S^3$. This is in 
agreement with what we found in the $M$ theory limit, since these become
fractional $M$2-branes. Thus the result of \cite{vafa} is consistent with
the results of this paper and strongly support the claim that $M$ theory
on $J$ is the strong coupling limit of the wrapped D6-brane system.
In fact, the amplitudes discussed 
in \cite{vafa} are well defined for all values
of the string coupling constant, including the strong coupling limit. 
Thus
they are also describing the physics of $M$ theory on $J$.

The previous statement is true regardless of the existence of a gravity dual
of the wrapped D6-brane system on $T^{*}({S^3})$. However, it would
be interesting to extend the `duality' between the wrapped D6-brane system
and $M$ theory on $J$ to the gravity dual of the wrapped D6-brane system
and another $M$ theory background. 
One way  of phrasing this desire
is to ask if there is an $M$ theory description of the IIA theory
on the resolved conifold with RR fluxes? We can proceed to answer this 
question as follows.

Consider the resolved conifold with only the 
$n$ units of RR 2-form flux turned on over the $S^2$. Because this
flux is non-trivial it implies that the $M$ theory circle is fibred
over the $S^2$ giving a circle bundle of first Chern number $n$. Thus
a natural proposal for the $M$ theory background is an $\mathbb{R^4}$-
bundle over ${S^3}/{\mathbb{Z_n}}$ - because the latter space indeed
is the total space of a degree $n$ bundle over $S^2$. Moreover,
at infinity this space looks just like infinity in $J$. We can now
propose a dual of the IIA theory on the resolved conifold with
only RR 2-form flux: $M$ theory on 
$\tilde{J}$ $\equiv$ $S({S^3}/{\mathbb{Z_n}})$ ie the standard
spin bundle of the Lens space. This space has a natural $G_2$-holonomy metric
\cite{bryssal} and in fact
$M$ theory on ${\tilde J}$ has the same symmetries as $M$ theory on $J$.

A crucial point of \cite{vafa} appeared to be
the requirement of RR 4-form and 6-form fluxes in the background. These
correspond to background values of the four-form field strength $G$ in $M$
theory on $\tilde{J}$.
The author is not aware of how to turn on $G$ in a $G_2$-holonomy background
without breaking supersymmetry. The arguments in favour of this have
always been in the context of compactifying to Minkowski space 
\cite{cr,b2}\ft{In \cite{cr} this is a supergravity calculation. In \cite{b2}
this is based upon a natural proposal for the superpotential of the
compactified theory \cite{guk}.}. 
However, it is plausible that this can be done
if the four dimensional spacetime normal to the seven manifold is anti de 
Sitter. This would be in accordance with the fact that the value of
the superpotential in the super Yang-Mills vacuum is non-zero and therefore
when embedded into a theory containing gravity contributes to a negative
cosmological constant in the vacuum. Our lack of understanding of this point
also reflects the fact that 
we essentially ignored gravity when
studying $M$ theory on $J$.

However, we have certainly identified what we consider to be
a good candidate for the $M$ theory background dual to IIA theory on
the resolved conifold with background RR 2-form fluxes.

The proposed model (and - if it exists - the model with $G$-flux)
seem to have some quite striking
features which would be required of a dual description of $M$ theory on $J$.
For instance the theory contains $\mathbb{Z_n}$-charged strings which arise
from wrapping the $M$2-brane around non-trivial
1-cycles which the generate of the fundamental group
of the Lens space. 
These are naturally identified with the super QCD strings.

The corresponding Type IIA model in three dimensions has BPS domain walls 
corresponding to D4-branes wrapping the Lens space. Since the fundamental
group of the Lens space is $\mathbb{Z_n}$ there are $n$ classical
vacua of the D4-brane world-volume theory corresponding to the $n$ inequivalent
irreducible representations of $\mathbb{Z_n}$. 
These are $n$ distinct domain walls and their spectrum - because they are BPS
states - is apparently the same in the quantum string theory \cite{gv}.

In the $M$ theory limit these correspond
to $M$5-branes wrapping the Lens space. 
These apparently have
non-trivial flat $B$-field connections on their world volumes.
If we now consider `reducing' $M$ theory on ${\tilde J}$ to 
give the IIA theory
on the resolved conifold with $n$ units of RR 2-form flux (as opposed
to considering IIA theory on ${\tilde J}$) then these $M$5-brane domain
walls go over to D$4$-brane domain walls which wrap the $S^2$. These 
D$4$-branes were identified in \cite{vafa} 
with the domain walls corresponding to
those expected in the super Yang-Mills theory.

One puzzling point however is that one only expects to see $n$ - $1$ distinct
domain walls in the super Yang-Mills theory, whereas here we appear to find
$n$. The remainder of this paragraph is devoted to some speculative
remarks about this issue.
In the Type IIA theory on ${\tilde J}$, $n$ - 1 of these walls is certainly
distinguished, since they carry non-zero $\mathbb{Z_n}$ quantum numbers. The
trivially charged domain wall on the other hand can be thought of as
a wrapped D4-brane on $S({S^3})$ which is $\mathbb{Z_n}$ invariant. In the
Type IIA theory on $S({S^3})$ this domain wall can be regarded as separating
flat three dimensional Minkowski space from three dimensional anti de Sitter
space or perhaps two anti de Sitter regions. $M$ theory on $S({S^3})$ is
the gravity dual of $M$ theory on $S({S^3})$! Thus, the additional wall in
the theory on ${\tilde J}$
corresponding to the affine node appears to map 
in the theory on $J$ to a domain wall separating
two regions of empty space in $M$ theory on $J$, and it is unnatural
to identify these with vacua of super Yang-Mills. Perhaps there is a more
rigorous argument which identifies the $n$ - 1 domain walls.

Most of the results of this paper can be generalised to theories with
$SO(2n)$ and $E_n$ symmetries by replacing $\mathbb{Z_n}$ with the $D$ and
$E$ type subgroups of $SU(2)$.

\bigskip
\large
\noindent
{\bf {\sf Acknowledgements.}}
\normalsize
\bigskip

The author would like to thank M. Douglas,
R. Gopakumar, G. Moore, A. Rajaraman
for recent discussions; J. Gauntlett, T. Hollowood
and D. Tong for discussions over
the past year and a half 
and J. Maldacena for informing us about \cite{malvaf}
prior to publication.

\end{document}